\documentclass[aps,showpacs]{revtex4}%
\usepackage{amssymb}
\usepackage{amsmath}
\usepackage{graphicx}
\usepackage{amsfonts}%
\begin{document}
\title{Reality and causality in quantum gravity modified electrodynamics }
\author{Santiago A. Mart\'\i nez$^{1}$, R. Montemayor$^{1}$ and Luis F. Urrutia$^{2}$}
\affiliation{$^{2}$ Instituto de Ciencias Nucleares, Universidad Nacional Aut{\'o}noma de
M{\'e}xico, A. Postal 70-543, 04510 M{\'e}xico D.F., M{\'e}xico}

\begin{abstract}
We present a general description of the propagation properties of quantum
gravity modified electrodynamics characterized by constitutive relations up to
second order in the correction parameter. The effective description
corresponds to an electrodynamics in a dispersive and absorptive non-local
medium, where the Green functions and the refraction indices can be explicitly
calculated. The reality of the electromagnetic field together with the
requirement of causal propagation in a given referrence frame leads to
restrictions in the form of such refraction indices. In particular, absorption
must be present in all cases and, contrary to the usual assumption, it is the
dominant aspect in those effective models which exhibit linear effects in the
correction parameter not related to birefringence. In such a situation
absorption is linear while propagation is quadratical in the correction parameter.

\end{abstract}
\maketitle

\section{Introduction}

The different effective descriptions of quantum gravity effects at low
energies can usually be expressed in terms of modified dispersion relations,
with a polynomial dependence on the energy and the momentum. Such
modifications include standard Lorentz invariance violations as well as
possible extensions of Lorentz covariance\cite{REVIEWS}. In the case of
electrodynamics these heuristic approaches can be interpreted in terms of an
effective field theory describing the propagation of the electromagnetic field
in an effective medium, provided that space-time coordinates and momenta
commute\cite{EDMED1,EDMED2,ELLIS2,ELLIS3,GRAVANIS,EDMED3,EDMED4}. Otherwise,
as it happens for example in double special relativity models, the ordering
ambiguity in the Fourier transform imposes a different approach outside the
scope of the present work\cite{SNYDER}. In fact, most of the models considered
in the literature, regardless of their quantum origin, are ultimately
interpreted as a standard low energy effective theory when their proponents
want to probe the corresponding phenomenological consequences by making use of
the standard meaning of the required observables, such as energy and momentum.
Of course, it is possible to abandon the effective field theory framework but
then it is mandatory to revise the standard concept of observables together
with unfolding the rules to measure them.

In the effective field theory version of quantum gravity inspired modified
electrodynamics the equations of motion acquire the structure of typical
Maxwell equations in a medium\cite{LANDAU1}
\begin{align}
&  i\mathbf{k}\cdot\mathbf{D}=4\pi\rho,\;\;\;\;\;\;\mathbf{k}\cdot
\mathbf{B}=0,\label{GENMAXW1}\\
&  \mathbf{k}\times\mathbf{E}-\omega\mathbf{B}=0,\;\;i\mathbf{k}%
\times\mathbf{H}+i\omega\mathbf{D}=4\pi\mathbf{j}, \label{GENMAXW2}%
\end{align}
where the $\mathbf{D}$ and $\mathbf{H}$ fields are characterized by
constitutive relations of the form%
\begin{align}
D^{i}  &  =\alpha^{ij}E_{j}+\rho^{ij}B_{j},\label{CRD}\\
H^{i}  &  =\beta^{ij}B_{j}+\sigma^{ij}E_{j}. \label{CRH}%
\end{align}
The coefficients $\alpha^{ij}$, $\rho^{ij}$, $\beta^{ij}$, and $\sigma^{ij}$
are polynomials in $\omega$ and $k^{i}$. Once these constitutive relations are
given, we can proceed as usual in the case of an electromagnetic field in a
non-homogeneous medium. From a heuristic point of view, it is interesting to
remark that these constitutive relations correspond to effective media that
are non-local in space and time, which can be interpreted as a footprint of
the granularity induced by quantum gravity. Although it seems that these media
could be in principle quite arbitrary, the requirements of reality of the
electromagnetic field together with causal propagation lead to constraints
that must be satisfied by these effective theories on very general grounds.

Our definition of causality at the effective theory level is the standard one,
whereby the field strengths are obtained from the sources via retarded Green
functions. It is important to remark that here we will deal only with the
aspects of causality related to the dispersive character of quantum gravity
effects as they are manifested in a given reference frame, a character related
to a generalized susceptibility theorem. There is another more geometrical
aspect related to Lorentz transformations connecting different reference
frames, associated to the possible existence of closed causal curves which
could appear when there are superluminal velocities. No matter how close to
$c$ the superluminal velocities are in a given reference frame, in a boosted
enough frame these particles will propagate to the past. The analysis of the
possible existence of closed causal curves will not be discussed here.

In relation to Eqs. (\ref{GENMAXW1}-\ref{GENMAXW2}), we adhere to the point of
view of Ref. \cite{HEHL} where $\mathbf{E}$, $\mathbf{B}$, $\mathbf{D}$, and
$\mathbf{H}$ are considered as fundamentals fields, even in vacuum. In fact,
the excitations $\mathbf{D}$ and $\mathbf{H}$ are the potentials for the
charge density $\rho$ and the current density ($\mathbf{j+\partial}%
_{t}\mathbf{D}$), respectively. Thus, they are directly related to charge
conservation. On the other hand the electric and magnetic field strenghts
$\mathbf{E}$ and $\mathbf{B}$ are the forces acting on unit charges. Our
notation does not reflect their intrisic geometrical content described by the
characterization of $\mathbf{D}$ and $\mathbf{B}$ as three-dimensional
two-forms, together with \ $\mathbf{E}$ and $\mathbf{H}$ as one-forms. It
refers to the corresponding components of the forms where the Hodge duality
has been used to rewrite antisymmetric two-form components as one-form components.

To introduce our effective field approach we will briefly recall the
Lagrangian for the electromagnetic field in a non-dispersive medium. It is
usually written as
\begin{equation}
L=-\frac{1}{16\pi}F_{\mu\nu}\chi^{\left[  \mu\nu\right]  \left[  \alpha
\beta\right]  }F_{\alpha\beta}-j_{\mu}A^{\mu}, \label{L}%
\end{equation}
where $F_{\mu\nu}=\partial_{\mu}A_{\nu}-\partial_{\nu}A_{\mu}$, and
$\chi^{\left[  \mu\nu\right]  \left[  \alpha\beta\right]  }$ is a constant
tensor that encodes the information about the medium. This structure, where
the basic dynamical field is $A_{\mu}$, warrants gauge invariance and hence
charge conservation. The dynamics is given by the equations of motion%
\begin{equation}
\partial_{\mu}H^{\mu\nu}=4\pi j^{\nu}, \label{me}%
\end{equation}
together with the constitutive relations%
\begin{equation}
H^{\mu\nu}(x)=\chi^{\left[  \mu\nu\right]  \left[  \alpha\beta\right]
}F_{\alpha\beta}(x).
\end{equation}
Defining, as usual, the electric and magnetic field strengths as $F_{0i}%
=E_{i}$ and $F_{ij}=-\epsilon_{ijk}B_{k}$ respectively, and the corresponding
components of the excitation $H^{\mu\nu}$, $H^{0i}=D^{i}$ and $H^{ij}%
=-\epsilon^{ijk}H^{k}$, the constitutive relations become%
\begin{align}
D^{i}  &  =2\chi^{\left[  0i\right]  \left[  0j\right]  }E_{j}-\chi^{\left[
0i\right]  \left[  mn\right]  }\epsilon_{mnj}B_{j},\label{CR1}\\
H^{i}  &  =\epsilon_{ilk}\chi^{\left[  lk\right]  \left[  0j\right]  }%
E_{j}-\frac{1}{2}\epsilon_{ilk}\chi^{\left[  lk\right]  \left[  mn\right]
}\epsilon_{mnj}B_{j}, \label{CR2}%
\end{align}
and the effective equations of motion acquire the usual form for an
electromagnetic field in a medium given in the relations (\ref{GENMAXW1}%
-\ref{GENMAXW2}).

In the case of a non-local space-time medium this formalism is also suitable,
but now we have%
\begin{equation}
L=-\frac{1}{16\pi}\int\;d^{4}\tilde{x}\;F_{\mu\nu}(x^{\sigma})\chi^{\left[
\mu\nu\right]  \left[  \alpha\beta\right]  }(x^{\sigma}-\tilde{x}^{\sigma
})\;F_{\alpha\beta}(\tilde{x}^{\sigma}), \label{lt}%
\end{equation}
instead of (\ref{L}). The constitutive relations become%
\[
H^{\mu\nu}(x)=\int d^{4}\tilde{x}\;\chi^{\left[  \mu\nu\right]  \left[
\alpha\beta\right]  }(x^{\sigma}-\tilde{x}^{\sigma})F_{\alpha\beta}(\tilde
{x}).
\]
Demanding that the Lagrangian (\ref{lt}) be real for real fields
$F_{\alpha\beta}$ ($\mathbf{E}$, $\mathbf{B}$) implies also the reality of
$H^{\mu\nu}$ ($\mathbf{D}$, $\mathbf{H}$). Writing $\chi^{\left[  \mu
\nu\right]  \left[  \alpha\beta\right]  }(x^{\sigma}-\tilde{x}^{\sigma})$ in
terms of its Fourier transform%
\begin{equation}
\chi^{\left[  \mu\nu\right]  \left[  \alpha\beta\right]  }(x^{\sigma}%
-\tilde{x}^{\sigma})=\int d^{4}k\;e^{-ik\cdot(x-\tilde{x})}\;\hat{\chi
}^{\left[  \mu\nu\right]  \left[  \alpha\beta\right]  }(k^{\sigma}),
\end{equation}
we can easily demonstrate that (\ref{lt}) can also be written%
\begin{equation}
L=-\frac{1}{16\pi}F_{\mu\nu}(x^{\sigma})\;\left(  {\hat{\chi}}^{\left[  \mu
\nu\right]  \left[  \alpha\beta\right]  }(i\partial_{\sigma})\;F_{\alpha\beta
}(x^{\sigma})\right) . \label{LAG}%
\end{equation}
This shows that for a non-local medium we can use a Lagrangian with the same
form as (\ref{L}), but now with ${\hat{\chi}}^{\left[  \mu\nu\right]  \left[
\alpha\beta\right]  }$ being a derivative operator. Thus we are dealing with a
higher order theory for the variables $A^{\mu}$. In particular this implies
that dynamical functions, such as the density of energy and momentum, will
depend not only on the fields $\mathbf{E}$, $\mathbf{B}$, $\mathbf{D}$, and
$\mathbf{H}$, but also on their space-time derivatives. This will pose
additional constraints regarding the causality of the system, since the
causality of a given function does not warrant the causality of its space-time
derivatives due to the presence of the energy cutoff in the effective theory
(see Appendix A).

If ${\hat{\chi}}^{\left[  \mu\nu\right]  \left[  \alpha\beta\right]  }$ is
symmetric, in the sense that for each set of index values ($\mu$, $\nu$,
$\alpha$, $\beta$) (no sum with respect to repeated indices)
\begin{equation}
\int d^{4}x\;F_{\mu\nu}\left(  {\hat{\chi}}^{\left[  \mu\nu\right]  \left[
\alpha\beta\right]  }F_{\alpha\beta}\right)  =\int d^{4}x\;F_{\alpha\beta
}\left(  {\hat{\chi}}^{\left[  \alpha\beta\right]  \left[  \mu\nu\right]
}F_{\mu\nu}\right)  , \label{sym}%
\end{equation}
is satisfied. Thus it is possible to perform integrations by parts rendering
the equations of motion of the same form as the usual non-operator case. In
the following we assume that this property holds. The above relation
(\ref{sym}) can be rewritten as%
\begin{equation}
{\hat{\chi}}^{\left[  \mu\nu\right]  \left[  \alpha\beta\right]  }%
(i\partial_{\sigma})={\hat{\chi}}^{\left[  \alpha\beta\right]  \left[  \mu
\nu\right]  }(-i\partial_{\sigma}),
\end{equation}

The expansion of the ${\hat{\chi}}^{\left[  \mu\nu\right]  \left[  \alpha
\beta\right]  }$ operator in terms of space-time components%
\begin{equation}
L=-\frac{1}{4\pi}F_{0i}{\hat{\chi}}^{\left[  0i\right]  \left[  0m\right]
}F_{0m}-\frac{1}{4\pi}F_{0i}{\hat{\chi}}^{\left[  0i\right]  \left[
mn\right]  }F_{mn}-\frac{1}{16\pi}F_{ij}{\hat{\chi}}^{\left[  ij\right]
\left[  mn\right]  }F_{mn}-j_{\mu}A^{\mu} \label{LIVL}%
\end{equation}
can also be used to characterize a quantum gravity modified electromagnetic
Lagrangian which preserves 3-d isotropy, provided that ${\hat{\chi}}^{\left[
0i\right]  \left[  0i\right]  }$, ${\hat{\chi}}^{\left[  0i\right]  \left[
jk\right]  }$, ${\hat{\chi}}^{\left[  jk\right]  \left[  0i\right]  }$,
${\hat{\chi}}^{\left[  ij\right]  \left[  mn\right]  }$ are tensors under
spatial rotations, but not the components of a ${\hat{\chi}}^{\left[  \mu
\nu\right]  \left[  \alpha\beta\right]  }$ standard Lorentz tensor.

In this paper we use this approach to discuss the main properties of quantum
gravity induced effects in electrodynamics, which are labelled by a parameter
${\tilde{\xi}}$ usually considered to be proportional to the inverse Planck
mass $M_{P}$. We provide a general description of such modified
electrodynamics including expressions for the Green functions as well as the
corresponding index of refractions, in terms of a parameterization of the
constitutive relations up to second order in ${\tilde{\xi}}$. Some preliminary
work along these directions have already been presented in Ref. \cite{PDELC}.
We also discuss the implications of reality and causality in the field
propagation, applying them to some specific cases already considered in the
literature. In a different context, causality in Lorentz invariance violating
theories has previously been discussed in Ref. \cite{LEHNERT}.

\section{Constitutive relations and equations of motion}

In the following we consider corrections up to second order in the parameter
${\tilde{\xi}}$, assuming for simplicity that we are in a reference frame
where there is invariance under rotations. This would correspond, for example,
to the rest frame $V^{\mu}=(1,\mathbf{0})$ in the Myers-Pospelov
model\cite{MP}. Our approximation includes contributions from dimension five
and six operators in the language of effective field theories. The modified
Lagrangian (\ref{LIVL}) yields the equations of motion:%
\begin{align}
\partial_{i}\left[  2{\hat{\chi}}^{\left[  i0\right]  \left[  0m\right]
}F_{0m}+{\hat{\chi}}^{\left[  i0\right]  \left[  mn\right]  }F_{mn}\right]
&  =4\pi j^{0},\label{EM1}\\
\partial_{j}\left[  2{\hat{\chi}}^{\left[  ji\right]  \left[  0n\right]
}F_{0n}+{\hat{\chi}}^{\left[  ji\right]  \left[  mn\right]  }F_{mn}\right]
-\partial_{0}\left[  2{\hat{\chi}}^{\left[  i0\right]  \left[  0m\right]
}F_{0m}+{\hat{\chi}}^{\left[  i0\right]  \left[  mn\right]  }F_{mn}\right]
&  =4\pi j^{i}, \label{EM2}%
\end{align}
and the structure of the field tensor $F_{\mu\nu}$ leads to%
\begin{equation}
\partial_{\sigma}F_{\mu\nu}+\partial_{\mu}F_{\nu\sigma}+\partial_{\nu
}F_{\sigma\mu}=0.
\end{equation}

Comparing Eqs. (\ref{CR1}-\ref{CR2}) and (\ref{CRD}-\ref{CRH}) we have%
\begin{align}
{\hat{\chi}}^{\left[  0i\right]  \left[  0j\right]  }  &  =\frac{1}{2}%
\alpha^{ij},\ \ \ \ \ \ \ \ \ \ \ \ \ \ \ \ \ \ \ {\hat{\chi}}^{\left[
0i\right]  \left[  mn\right]  }=-\frac{1}{2}\epsilon_{mnj}\rho^{ij}%
,\label{chi1}\\
{\hat{\chi}}^{\left[  mn\right]  \left[  0j\right]  }  &  =+\frac{1}%
{2}\epsilon_{mni}\sigma^{ij},\ \ \ \ \ \ \ \ \ {\hat{\chi}}^{\left[
lk\right]  \left[  mn\right]  }=+\frac{1}{2}\epsilon_{kli}\beta^{ij}%
\epsilon_{jmn}. \label{chi2}%
\end{align}
In order to analyze the propagation of the fields and to define a refraction
index it is important to make explicit the dependence of the constitutive
relations on the frequency $\omega$ and the momentum $\mathbf{k}$. To achieve
this we expand their coefficients in space derivatives. Taking into account
that these models can be understood as perturbative descriptions in terms of
the parameter ${\tilde{\xi}}=\xi/M_{P}$ we have, up to order ${\tilde{\xi}%
}^{2}$,
\begin{align}
\alpha^{ij}  &  =\alpha_{0}(\partial_{t})\eta^{ij}+\alpha_{1}(\partial
_{t}){\tilde{\xi}}\epsilon^{ijr}\partial_{r}+\alpha_{2}(\partial_{t}%
){\tilde{\xi}}^{2}\partial^{i}\partial^{j},\label{alpha}\\
\rho^{ij}  &  =\rho_{0}(\partial_{t})\eta^{ij}+\rho_{1}(\partial_{t}%
){\tilde{\xi}}\epsilon^{ijr}\partial_{r}+\rho_{2}(\partial_{t}){\tilde{\xi}%
}^{2}\partial^{i}\partial^{j},\\
\sigma^{ij}  &  =\sigma_{0}(\partial_{t})\eta^{ij}+\sigma_{1}(\partial
_{t}){\tilde{\xi}}\epsilon^{ijr}\partial_{r}+\sigma_{2}(\partial_{t}%
){\tilde{\xi}}^{2}\partial^{i}\partial^{j},\\
\beta^{ij}  &  =\beta_{0}(\partial_{t})\eta^{ij}+\beta_{1}(\partial
_{t}){\tilde{\xi}}\epsilon^{ijr}\partial_{r}+\beta_{2}(\partial_{t}%
){\tilde{\xi}}^{2}\partial^{i}\partial^{j}, \label{rho}%
\end{align}
where $\alpha_{A}$, $\beta_{A}$, $\sigma_{A}$, and $\rho_{A}$, with $A=0,1,2$,
are $S0(3)$ scalar operators, according to the assumption that there is no
preferred spatial direction. By relaxing this assumption it is possible to
generalize the present approach to models that exhibit some preferred spatial
directions. Examples of these last models are given in Refs. \cite{MP},
\cite{gamboa}, and \cite{balley}, where a Lorentz violating vector $n^{\mu
}=(n^{0},n^{i})$ appears. Only in the reference frame where $n^{i}=0$ do they
yield an isotropic propagation, otherwise anisotropic effects occur. For
simplicity, in the following we restrict our discussion to the isotropic case.

The symmetry of ${\hat{\chi}}$ in Eq. (\ref{sym}) implies that the terms in
$\alpha^{ij}$ and $\beta^{ij}$ which contain an even number of derivatives are
symmetric under $i\leftrightarrow j$, while the terms with an odd number of
derivatives are antisymmetric. In the case of $\rho^{ij}$ and $\sigma^{ij}$
Eq. (\ref{sym}) leads to%
\begin{equation}
\rho_{i}=-\sigma_{i}.
\end{equation}

The property $\mathbf{\nabla}\mathbf{\cdot B}=0$ implies that $\beta_{2}$ and
$\rho_{2}$ are irrelevant, and thus we take $\beta_{2}=\rho_{2}=0$ (and hence
$\sigma_{2}=0$). Thus Eqs. (\ref{alpha}-\ref{rho}) finally reduce to%
\begin{align}
\alpha^{ij}  &  =\alpha_{0}\eta^{ij}+\alpha_{1}{\tilde{\xi}}\epsilon
^{ijr}\partial_{r}+\alpha_{2}{\tilde{\xi}}^{2}\partial^{i}\partial
^{j},\label{CRFI}\\
\rho^{ij}  &  =-\sigma^{ij}=-\sigma_{0}(t)\eta^{ij}-\sigma_{1}{\tilde{\xi}%
}\epsilon^{ijr}\partial_{r},\\
\beta^{ij}  &  =\beta_{0}\eta^{ij}+\beta_{1}{\tilde{\xi}}\epsilon
^{ijr}\partial_{r}, \label{CRFIN}%
\end{align}
which, together with the first of Eqs. (\ref{GENMAXW2}), yield the following
constitutive relations in momentum space%
\begin{align}
\mathbf{D}  &  =\left(  \alpha_{0}+\alpha_{2}{\mathbf{k}^{2}\tilde{\xi}}^{2}
\right)  \mathbf{E}-\left(  \sigma_{0}+i\alpha_{1}{\omega\tilde{\xi} }\right)
\mathbf{B}+\left(  i\sigma_{1}{\tilde{\xi}}+\alpha_{2}\omega{\tilde{\xi}}%
^{2}\right)  \left(  \mathbf{k\times B}\right)  ,\label{D1}\\
\mathbf{H}  &  =\left(  \beta_{0}-i{\sigma_{1}\omega}{\tilde{\xi}}\right)
\mathbf{B}-i\beta_{1}{\tilde{\xi}}\left(  \mathbf{k\times B}\right)
+\sigma_{0}\mathbf{E}. \label{H1}%
\end{align}
In the approximation to order ${\tilde{\xi}}^{2}$ here considered, we have
${\tilde{\xi}}^{2}k^{2}\simeq{\tilde{\xi}}^{2}\omega^{2}$ and thus we can
write%
\begin{align}
\mathbf{D}  &  =d_{1}(\omega)\mathbf{E}+id_{2}(\omega)\mathbf{B}+d_{3}%
(\omega){\tilde{\xi}}\left(  \mathbf{k\times B}\right)  ,\label{D2}\\
\mathbf{H}  &  =h_{1}(\omega)\mathbf{B}+i{h}_{2}(\omega)\mathbf{E}%
+ih_{3}(\omega){\tilde{\xi}}\left(  \mathbf{k\times B}\right)  , \label{H2}%
\end{align}
where the functions $d_{i}(\omega)$ and $h_{i}(\omega)$ depend only on
$\omega$ and admit a series expansion in powers of ${\tilde{\xi}}\omega$,
characterizing each specific model. From Eqs. (\ref{GENMAXW1}-\ref{GENMAXW2})
we get the equations for $\mathbf{E}$ and $\mathbf{B}$%
\begin{align}
id_{1}\left(  \mathbf{k}\cdot\mathbf{E}\right)   &  =4\pi\rho\left(
\omega,\mathbf{k}\right)  ,\label{INHOM1}\\
i\omega d_{1}\mathbf{E}+\left(  h_{3}k^{2}-g(\omega)\right)  {\tilde{\xi}%
}\mathbf{B}+\left(  \omega d_{3}{\tilde{\xi}+}h_{1}\right)  \left(
i\mathbf{k}\times\mathbf{B}\right)   &  =4\pi\mathbf{j}\left(  \omega
,\mathbf{k}\right)  , \label{INHOM2}%
\end{align}
where we denote%
\begin{equation}
\left(  d_{2}+h_{2}\right)  \omega=g(\omega){\tilde{\xi}}.
\end{equation}
The expressions (\ref{D1}-\ref{H1}) indeed indicate that the above combination
is of order ${\tilde{\xi}}$. Thus we see that in fact there are only three
independent functions of $\omega$ and $k$ which determine the dynamics, which
are%
\begin{equation}
P=d_{1},\quad Q=h_{1}+\omega d_{3}{\tilde{\xi}},\quad R=\left(  h_{3}%
k^{2}-g(\omega)\right)  {\tilde{\xi}}. \label{R}%
\end{equation}
Using the homogeneous equation $\omega\mathbf{B}=\mathbf{k}\times\mathbf{E} $,
which yields $\omega\left(  \mathbf{k}\times\mathbf{B}\right)  =\left(
\mathbf{k\cdot E}\right)  \mathbf{k-}k^{2}\mathbf{E}$, and charge
conservation, $\omega\rho-\mathbf{k\cdot J}=0$, we can decouple the equations
for the fields $\mathbf{E}$ and $\mathbf{B}$ obtaining
\begin{align}
\,\mathbf{k\cdot E}  &  =\frac{4\pi}{i\omega P}\left(  \mathbf{k\cdot
j}\right)  ,\label{CCINHOM1}\\
i\left(  \omega^{2}P-k^{2}Q\right)  \mathbf{E}+R\,\left(  \mathbf{k}%
\times\mathbf{E}\right)   &  =4\pi\omega\left[  \mathbf{j-}\frac{k^{2}%
Q}{\omega^{2}P}\left(  \mathbf{\hat{k}\cdot j}\right)  \mathbf{\hat{k}%
}\right]  , \label{CCINHOM2}%
\end{align}
and%
\begin{align}
\,\mathbf{k\cdot B}  &  =0,\\
i\left(  \omega^{2}P-k^{2}Q\right)  \mathbf{B}+R\,\left(  \mathbf{k}%
\times\mathbf{B}\right)   &  =4\pi k\;\mathbf{\hat{k}}\times\mathbf{j}.
\label{B}%
\end{align}

Finally, we can introduce the standard potentials $\Phi$ and $\mathbf{A}$
\begin{equation}
\mathbf{B}=i\mathbf{k}\times\mathbf{A},\qquad\mathbf{E}=i\omega\mathbf{A-}%
i\mathbf{k}\Phi.
\end{equation}
As in the usual case we can use the radiation gauge, $\mathbf{k}%
\cdot\mathbf{A=}0$, in which case we have
\begin{align}
\Phi &  =4\pi\left(  k^{2}P\right)  ^{-1}\rho,\label{PHIRG}\\
\left(  k^{2}Q-\omega^{2}P\right)  \mathbf{A}+R\;\left(  i\mathbf{k}%
\times\mathbf{A}\right)   &  =4\pi\left[  \mathbf{j-}\left(  \mathbf{j}%
\cdot\mathbf{\hat{k}}\right)  \mathbf{\hat{k}}\right]  =4\pi\mathbf{j}_{T},
\label{EQARG}%
\end{align}
from Eqs. (\ref{INHOM1}-\ref{INHOM2}). The presence of birefringence depends
on the parity violating term proportional to $kR$, where we have a crucial
factor of $i$. This makes it clear that a possible diagonalization can be
obtained by using a complex basis, which is precisely the circular
polarization basis. In fact, decomposing the vector potential and the current
in such a basis%
\begin{equation}
\mathbf{A}=\mathbf{A}^{+}+\mathbf{A}^{-},\;\;\;\;\mathbf{j}_{T}=\mathbf{j}%
_{T}^{+}+\mathbf{j}_{T}^{-},
\end{equation}
and recalling the basic properties $\mathbf{\hat{k}}\times\mathbf{A}%
^{+}=-i\mathbf{A}^{+}$, $\mathbf{\hat{k}}\times\mathbf{A}^{-}=i\mathbf{A}^{-}%
$, we can separate (\ref{EQARG}) into the uncoupled equations \cite{PDELC}
\begin{equation}
\left[  k^{2}Q-\omega^{2}P+\lambda k\,R\right]  \mathbf{A}^{\lambda}%
=4\pi\mathbf{j}_{T}^{\lambda},\;\;\;\;\lambda=\pm1. \label{af}%
\end{equation}
In terms of the basic functions introduced in the constitutive relations
(\ref{D2}-\ref{H2}), the factor in (\ref{af}) is rewritten as
\begin{equation}
k^{2}Q-\omega^{2}P+\lambda k\,R=\lambda h_{3}k^{3}{\tilde{\xi}+}\left(
h_{1}+d_{3}{\omega\tilde{\xi}}\right)  k^{2}-\lambda g{\tilde{\xi}}%
k-d_{1}\omega^{2}.
\end{equation}
This is the key expression to obtain the Green functions and the refraction indices.

\section{The energy-momentum density}

In order to derive the Poynting theorem for this generalized electrodynamics
we still need to make explicit the remaining dependence upon $\partial_{t}$ of
the operators $\alpha_{0}$, $\beta_{0}$, $\sigma_{0}$, $\alpha_{1}$,
$\beta_{1}$, $\sigma_{1}$, and $\alpha_{2}$, in Eqs. (\ref{CRFI}-\ref{CRFIN}).
We denote them generically by $\zeta_{0}$, $\zeta_{1}$ and $\zeta_{2}$,
respectively. Following the symmetry conditions imposed by Eq.(\ref{sym}) we
have, up to second order in ${\tilde{\xi}}$,%
\begin{equation}
\zeta_{0}=\zeta_{00}+\zeta_{02}{\tilde{\xi}}^{2}\partial_{t}^{2},\;\;\zeta
_{1}=\zeta_{10},\;\;\zeta_{2}=\zeta_{20}, \label{zeta}%
\end{equation}
where $\zeta_{00}$, $\zeta_{02}$, $\zeta_{10}$ and $\zeta_{20}$ are now
constant coefficients. Here we take
\begin{equation}
\alpha_{00}=\beta_{00}=1,\;\sigma_{00}=0,
\end{equation}
to recover the usual vacuum as the background for ${\tilde{\xi}=0}$. A long
but straightforward calculation starting from Maxwell equations
(\ref{GENMAXW1}-\ref{GENMAXW2}) in coordinate space leads to energy
conservation in the form%
\begin{equation}
\partial_{t}u+\mathbf{\nabla\cdot S}=-\mathbf{j\cdot E}. \label{CF}%
\end{equation}
Outside the sources, where we have the additional property ${\tilde{\xi}}%
^{2}\mathbf{\nabla\cdot E}=0$, the corresponding energy density $u$ and
Poynting vector $\mathbf{S}$ are%
\begin{align}
u  &  =\frac{1}{2}\left(  \mathbf{E}^{2}+\mathbf{B}^{2}\right)  -\tilde{\xi
}\frac{1}{2}\alpha_{10}\mathbf{E}\cdot\nabla\times\mathbf{E}\;-\tilde{\xi
}\frac{1}{2}\beta_{10}\mathbf{B}\cdot\nabla\times\mathbf{B}\nonumber\\
&  +\sigma_{02}\tilde{\xi}^{2}\left[  \left(  \partial_{t}\mathbf{E}\right)
\cdot\left(  \partial_{t}\mathbf{B}\right)  -\mathbf{E\cdot}\partial_{t}%
^{2}\mathbf{B}\right]  +\beta_{02}\frac{1}{2}\tilde{\xi}^{2}\left(
\partial_{t}\mathbf{B}\right)  ^{2}+\alpha_{02}\tilde{\xi}^{2}\left[
\mathbf{E}\cdot\partial_{t}^{2}\mathbf{E}-\frac{1}{2}\left(  \partial
_{t}\mathbf{E}\right)  ^{2}\right]  , \label{DENSEN}%
\end{align}%
\begin{align}
\mathbf{S}  &  =\mathbf{E}\times\mathbf{B}\;+\frac{1}{2}\tilde{\xi}\alpha
_{10}\mathbf{E}\times\partial_{t}\mathbf{E}\nonumber\\
&  \mathbf{+}\tilde{\xi}\beta_{10}\left[  \frac{1}{2}\partial_{t}%
\mathbf{B}\times\mathbf{B}-\mathbf{E}\times\left(  \mathbf{\nabla}%
\times\mathbf{B}\right)  \right]  +\sigma_{02}\tilde{\xi}^{2}\mathbf{E}%
\times\partial_{t}^{2}\mathbf{E}+\beta_{02}\tilde{\xi}^{2}\mathbf{E}%
\times\partial_{t}^{2}\mathbf{B.} \label{FLUXEN}%
\end{align}
In our case neither the energy density, nor the Poynting vector retain their
standard forms $(\mathbf{E\cdot D+B\cdot H})/2$ and $\mathbf{E\times H}$
respectively, because we are dealing with a higher order theory. This can be
seen from the application of the apropriately extended Noether theorem to the
Lagrangian density (\ref{LAG}), which includes up to third derivatives in the
photon field $A_{\mu}$. To this end we introduce
\begin{equation}
\hat{\chi}^{[\mu\nu][\alpha\beta]}=\chi_{0}^{[\mu\nu][\alpha\beta]}+\chi
_{1}^{[\mu\nu]\theta\lbrack\alpha\beta]}\partial_{\theta}+\chi_{2}^{[\mu
\nu]\{\theta\psi\}[\alpha\beta]}\partial_{\theta}\partial_{\psi},
\label{GENSUS}%
\end{equation}
as a more compact way of writting the parameterization (\ref{chi1}%
-\ref{chi2}), (\ref{alpha}-\ref{rho}) and (\ref{zeta}) of the generalized
susceptibility operator. The coefficients $\chi_{A}^{[\mu\nu]\{...\}[\alpha
\beta]}$, $A=0,1,2$, are constants and the gauge invariant energy momentum
tensor is\ \cite{MUOBREGON}
\begin{align}
T_{\sigma}^{\tau}  &  =-\delta_{\sigma}^{\tau}\;L-F_{\sigma\alpha}%
H^{\tau\alpha}-\frac{1}{4}\left(  \partial_{\sigma}F_{\mu\alpha}\right)
\chi_{1}^{[\theta\psi]\tau[\mu\alpha]}F_{\theta\psi}\nonumber\\
&  +\frac{1}{4}\left(  \partial_{\sigma}F_{\mu\alpha}\right)  \chi
_{2}^{[\theta\psi]\{\rho\tau\}[\mu\alpha]}\left(  \partial_{\rho}F_{\theta
\psi}\right)  -\frac{1}{4}F_{\theta\psi}\chi_{2}^{[\theta\psi]\{\nu\tau
\}[\mu\alpha] }\left(  \partial_{\sigma}\partial_{\nu}F_{\mu\alpha}\right)  .
\label{FINEMT}%
\end{align}
This satisfies $\partial_{\tau}T_{\sigma}^{\tau}=0$ outside the sources, in
virtue of Maxwell equations (\ref{me}). The standard contributions arise
precisely from the first two terms in the right hand side of (\ref{FINEMT}).
In fact we have%
\begin{align}
-L-F_{0\alpha}H^{0\alpha}  &  =\frac{1}{2}\left(  \mathbf{E\cdot D+B\cdot
H}\right)  ,\\
-F_{0\alpha}H^{i\alpha}  &  =\mathbf{E\times H}.
\end{align}
It is convenient to remember that these models have to be interpreted as
effective theories, in which the dynamics of quantum gravity fluctuations are
partially incorporated via the modified constitutive relations (\ref{CRD}%
-\ref{CRH}). In this sense they are anologous to the Euler-Heisenberg
effective Lagrangian for the electromagnetic field including quantum
corrections through non-linear terms, or to the purely classical Lorentz-Dirac
dynamics for a charge including the radiated field through a third order
derivative term. Thus the density of energy and momentum given by the Noether
theorem, $u$ and $\mathbf{S}$ of Eqs. (\ref{DENSEN}-\ref{FLUXEN}) in
coordinate space, contain in principle the contributions of the
electromagnetic field plus the quantum gravity fluctuations. These magnitudes
satisfy the continuity equation (\ref{CF}). Hence , if we consider a volume
$V$ outside the sources such that the total flux of $\mathbf{S}$ in its
boundary is null, it results that%
\begin{equation}
U=\int_{V}d^{3}x\ u
\end{equation}
is a constant of motion despite damping factors appearing in a plane wave
field. The density $u$ does not correspond to the usual expression in
electrodynamics, $(E^{2}+B^{2})/2$, which does not lead here to a constant of
motion. In terms of the analogy with an electromagnetic field propagating in a
medium, this can be interpreted as the typical effect of an active medium that
interchanges energy with the field. If this active medium absorbs or cedes
energy to the electromagnetic field it is a matter of the second law of
thermodynamics, which is not explicitely implemented in these models.

We will demand that the physical quantities $u$ and $\mathbf{S}$ be causally
related to the sources. The fields $\mathbf{E}$ and $\mathbf{B}$ are causal by
their construction in terms of retarded Green functions. Since we are dealing
with an effective theory with an energy cutoff, neither the time nor the
spatial derivatives of a causal function are causal, as discussed in Appendix
A. In this way, to meet the above requirement we have to impose causality, via
the Kramers-Kroning relations, to a sequence of derivatives of $\mathbf{E}$
and $\mathbf{B}$ dictated by the order in $\tilde{\xi}$ to which we want the
theory to hold. We have explored this situation only to first order in
$\tilde{\xi}$, where we can show that it is enough to demand also that
$\mathbf{D}$ and $\mathbf{H}$ be causal in order to guarantee the causality of
the energy density and the Poynting vector. Keeping the approximation to first
order we can use the zeroth order relations%
\begin{align}
\mathbf{\nabla\cdot E}  &  =0,\;\;\;\mathbf{\nabla\cdot B}=0,\\
\mathbf{\nabla\times E}  &  =-\partial_{t}\mathbf{B;\;\;\nabla\times
B}=\partial_{t}\mathbf{E,}%
\end{align}
in all derivative terms containing a factor $\tilde{\xi}$. In this way we can
rewrite $u$ and $\mathbf{S}$ only in terms of $\mathbf{\nabla\times E}$ and
$\mathbf{\nabla\times B}$. On the other hand, from the constitutive relations%
\begin{equation}
\mathbf{D}=\mathbf{E}+\sigma_{10}\tilde{\xi}\mathbf{\nabla}\times
\mathbf{B}-\alpha_{10}\tilde{\xi}\mathbf{\nabla}\times\mathbf{E},
\end{equation}%
\begin{equation}
\mathbf{H}=\mathbf{B}-\beta_{10}\tilde{\xi}\mathbf{\nabla}\times
\mathbf{B}-\sigma_{10}\tilde{\xi}\mathbf{\nabla}\times\mathbf{E}%
\end{equation}
and assuming $\alpha_{10}\beta_{10}+\left(  \sigma_{10}\right)  ^{2}\neq0$, we
can express the energy density and Poynting vector only in terms of
$\mathbf{E}$, $\mathbf{B}$, $\mathbf{D}$ and $\mathbf{H}$. The causality
requirement for $\mathbf{D}$ and $\mathbf{H}$ are explored in the next section.

\section{Reality and causality constraints}

The reality constraints are straightforward. From the constitutive relations
(\ref{D2}-\ref{H2}) the fields $\mathbf{D}$ and $\mathbf{H}$ are real for
$\mathbf{E}$ and $\mathbf{B}$ real if
\begin{align}
d_{1}^{\ast}(\omega)  &  =d_{1}(-\omega),\;\;\;d_{2}^{\ast}(\omega
)=-d_{2}(-\omega)\;,\;\;d_{3}^{\ast}(\omega)=-d_{3}(-\omega),\label{RC1}\\
h_{1}^{\ast}(\omega)  &  =h_{1}(-\omega),\;\;\;h_{2}^{\ast}(\omega
)=-h_{2}(-\omega),\;\;\;h_{3}^{\ast}(\omega)=h_{3}(-\omega). \label{RC2}%
\end{align}
Even though we will subsequently extend the variable $\omega$ to the complex
plane, the relations (\ref{RC1}-\ref{RC2}) are written for real $\omega$,
which represents the physical frequency. Nevertheless, in the case of an
absorptive background with a real frequency, the wave vector $\mathbf{k(}%
\omega\mathbf{)}$, and hence $k(\omega)=\sqrt{\mathbf{k}\cdot\mathbf{k}}$
together with the index of refraction $n(\omega)=k(\omega)/\omega$, will be
complex in general \cite{LANDAU1}. Unless explicitly stated, all our
subsequent expressions involving the frequency are written for real $\omega$.

To have a causal model it is necessary that not only the propagation of the
vector potential $\mathbf{A}$, together with $\mathbf{E}$ and $\mathbf{B}$, is
causal, but also the response of the medium coded in the constitutive
relations. In the following we discuss this last point, finding the causal
constraints which must satisfy the functions that appear in the constitutive
equations. To this end we generically denote the functions $d_{1}(\omega)$,
$id_{2}(\omega)$, $\omega\xi d_{3}(\omega)$, $h_{1}(\omega)$, $ih_{2}(\omega
)$, and $i\omega\xi h_{3}(\omega)$ by $\vartheta(\omega)$. In this way the
reality conditions are%
\begin{equation}
\mathrm{Re}\vartheta(\omega)=\mathrm{Re}\vartheta(-\omega
),\ \ \ \ \ \ \ \ \ \mathrm{Im}\vartheta(\omega)=-\mathrm{Im}\vartheta
(-\omega).
\end{equation}
A general discussion of causality and dispersion relations is given in
Ref.\cite{TOLL}. In brief, the causal response of the effective medium depends
on the absence of poles of $\vartheta(\omega)=\mathrm{Re}\,\vartheta
(\omega)+i\mathrm{Im}\,\vartheta(\omega)$ in the upper half complex plane
$\omega$. In addition, here it is necessary to take into account that we are
discussing effective models which lose their physical meaning at high
frequencies. Therefore, we must consider a finite range of frequencies instead
of the usual infinite one, and thus causality is characterized by the
generalized susceptibility theorem \cite{LANDAUDR}\cite{LANDAU1}
\begin{align}
\mathrm{Re}\,\vartheta(\omega)-\kappa &  =\frac{1}{\pi}P\int_{-\Omega}%
^{\Omega}d\omega^{\prime}\ \frac{\mathrm{Im}\,\vartheta(\omega^{\prime}%
)}{\omega^{\prime}-\omega}=\frac{2}{\pi}P\int_{0}^{\Omega}d\omega^{\prime
}\ \frac{\omega^{\prime}\,\mathrm{Im}\,\vartheta(\omega^{\prime})}%
{\omega^{\prime2}-\omega^{2}},\label{pp1}\\
\mathrm{Im}\,\vartheta(\omega) &  =-\frac{1}{\pi}P\int_{-\Omega}^{\Omega
}d\omega^{\prime}\ \frac{\mathrm{Re}\,\vartheta(\omega^{\prime})-\kappa
}{\omega^{\prime}-\omega}+\frac{A}{\omega},\label{pp2}%
\end{align}
where the cutoff $\Omega$ defines the region where the effective theory holds,
$|\omega|\lesssim\Omega$, $\kappa=\lim_{\omega\rightarrow\Omega}%
\mathrm{{Re}\,\vartheta(\omega)}$ and $A=-i\lim_{\omega\rightarrow0}\left[
\omega\vartheta(\omega)\right]  $. We assume that the contribution of the path
closing the loop in the upper half complex plane $\omega$ is negligible. We
have evaluated the above dispersion relations taking an inverse power law
behavior of $\vartheta(\omega)$ for $|\omega|\,>\,\Omega$ and matching the
functions at $\omega=\pm\Omega$, obtaining response coefficients which are of
the same order of magnitude as those in the proposed simpler approximation.
Via the dispersion relations, the introduction of a cutoff in $\omega$ leads
to corresponding cutoff in $k(\omega)$. This cutoff induces an effective
granularity in space-time at small distances, and can be readily seen as
follows. Lets us consider a given function in the $k$-space, $f(k)$. If there
is no cutoff in this space we have the corresponding Fourier transform to the
coordinate space%
\begin{equation}
f(x)=\frac{1}{2\pi}\int_{-\infty}^{\infty}dk\ e^{-ikx}f(k).\label{nct}%
\end{equation}
But if there is a cutoff $\chi$ in the $k$-space , i.e. $-\chi<k<\chi$,
instead of this we have%
\begin{equation}
\hat{f}(x)=\int_{-\chi}^{\chi}dk\ e^{ikx}\tilde{f}(k),\label{ct}%
\end{equation}
which, using Eq. (\ref{nct}), becomes related to $f(x)$ by an integral
transform%
\begin{equation}
\hat{f}(x)=\int_{-\infty}^{\infty}dx^{\prime}f(x^{\prime})\frac{\sin\left(
x-x^{\prime}\right)  \chi}{\pi\left(  x-x^{\prime}\right)  }.\label{nctct}%
\end{equation}
In the integrand of the right hand side of Eq. (\ref{nctct}) there now appears
a smeared distribution around $x$, with a width $\chi^{-1}$, which can be
interpreted as an effective granularity in the coordinate space. This
space-time granularity is telling us that we can not probe space-time at
distances smaller than $1/\chi$\ through the effective theory.

If we restrict ourselves to perturbative expressions, we can establish the
causal structure of the coefficients in terms of an expansion in $\tilde{\xi}
$. To be precise, let us write the real and imaginary parts of $\vartheta$ as%
\begin{align}
\mathrm{Re}\,\vartheta\left(  \omega\right)   &  =\theta_{0}+\theta_{2}%
\omega^{2}+\theta_{4}\omega^{4}+\mathcal{O}(\theta_{6}{\omega}^{6}),\\
\mathrm{Im}\,\vartheta\left(  \omega\right)   &  =\theta_{1}\omega+\theta
_{3}\omega^{3}+\mathcal{O}(\theta_{5}{\omega}^{5}).
\end{align}
In terms of the scale $\tilde{\xi}$ the notation is $\theta_{k}=\vartheta
_{k}\tilde{\xi}^{k}$. Moreover, $\vartheta(\omega)$ has no pole at $\omega=0$,
hence $A=0$, and we have $\mathrm{Re}\,\vartheta\left(  \Omega\right)
=\kappa$ at $\omega=\Omega$. Thus we get
\begin{equation}
\mathrm{Im}\,\vartheta(\omega)\simeq\frac{4}{\pi}\left[  -\left(  \theta
_{2}\Omega^{2}+\frac{2}{3}\theta_{4}\Omega^{4}\right)  \left(  \frac{\omega
}{\Omega}\right)  +\left(  \frac{1}{3}\theta_{2}\Omega^{2}-\frac{2}{3}%
\theta_{4}\Omega^{4}\right)  \left(  \frac{\omega}{\Omega}\right)
^{3}\right]  .
\end{equation}
This allows us to identify%
\begin{equation}
\theta_{1}=-\frac{4}{\pi\Omega}\left(  \theta_{2}\Omega^{2}+\frac{2}{3}%
\theta_{4}\Omega^{4}\right)  ,
\end{equation}
and solving for $\theta_{4}$ we can predict the third order contribution to
$\mathrm{Im}\,\vartheta(\omega)$ to be%
\begin{equation}
\theta_{3}=\frac{1}{\Omega^{2}}\left(  \theta_{1}+\frac{16}{3\pi}\theta
_{2}\Omega\right)  .
\end{equation}
Thus the final expression up to third order in $\tilde{\xi}$ is%
\begin{equation}
\vartheta(\omega)\simeq\vartheta_{0}+i\vartheta_{1}\omega\tilde{\xi}%
+\vartheta_{2}\omega^{2}\tilde{\xi}^{2}+\frac{i}{\Omega\tilde{\xi}}\left(
\frac{\vartheta_{1}}{\Omega\tilde{\xi}}+\frac{16\vartheta_{2}}{3\pi}\right)
\omega^{3}\tilde{\xi}^{3}.
\end{equation}
This is the structure imposed by causality on every coefficient $\vartheta$ in
the relation of $\mathbf{E}$, $\mathbf{B}$ with $\mathbf{D}$, $\mathbf{H}$.

\section{Green functions in the radiation gauge}

The retarded Green function for the potential $\mathbf{A}$ and the field
strengths $\mathbf{E}$, $\mathbf{B}$ (see Eqs. (\ref{CCINHOM2}-\ref{B}%
-\ref{EQARG})) can be written in terms of the circular polarization basis
\begin{equation}
G_{ij}^{ret}(\omega,\mathbf{r})=\int\frac{d^{3}k}{\left(  2\pi\right)  ^{3}%
}e^{i\mathbf{k}\cdot\mathbf{r}}\,\tilde{G}_{ij}^{ret}(\omega,\mathbf{k}%
)=\frac{1}{2}\int\frac{d^{3}k}{\left(  2\pi\right)  ^{3}}e^{i\mathbf{k}%
\cdot\mathbf{r}}\,\sum_{\lambda}G^{\lambda}(\omega,\mathbf{k})\left(
\delta_{ik}\mathbf{-}\frac{{k}_{i}{k}_{k}}{k^{2}}+i\lambda\epsilon_{irk}%
\frac{{k}_{r}}{k}\right)  , \label{gg}%
\end{equation}
where ${\hat{k}}_{i}=k_{i}/|\mathbf{k}|$, $k=|\mathbf{k}|$, and $G^{\lambda
}(\omega,\mathbf{k})$ is obtained from Eq.(\ref{af}),%
\begin{equation}
G^{\lambda}(\omega,\mathbf{k})=\frac{1}{k^{2}Q-\omega^{2}P+\lambda
k\,R},\;\;\;\;\lambda=\pm1. \label{gl}%
\end{equation}
To obtain the causal Green functions the analytic continuation $\omega
\rightarrow\omega+i\epsilon$ is taken. Only the poles in the upper half plane
of the variable $k$ make a contribution to the integration. The first step is
to rewrite the denominator in Eq. (\ref{gl}) in a convenient form. This is
done by successive rescalings leading to%
\begin{equation}
Qk^{2}-P\omega^{2}+\lambda kR=-\frac{1}{3}n_{0}^{2}\omega^{2}Qa(M^{\lambda
}-M_{0})(M^{\lambda}-M_{+})(M^{\lambda}-M_{-}), \label{gl1}%
\end{equation}
where we introduce the notation
\begin{align}
Q  &  =h_{1}+d_{3}{\omega\tilde{\xi}},\ \ \;\;\chi=\tilde{\xi}\omega
Q^{-1},\ \ \;\;n_{0}^{2}=d_{1}Q^{-1},\nonumber\\
&  \ \ \ \ \ a=3h_{3}n_{0}\chi,\;\ \ \ \ \ \;M^{\lambda}=\lambda k\left(
n_{0}\omega\right)  ^{-1}.
\end{align}
The corresponding roots are given by
\begin{align}
M_{0}  &  =\frac{1}{a}+2^{1/3}\frac{1+ac}{aA}+\frac{A}{2^{1/3}a},\nonumber\\
M_{-}  &  =\frac{1}{a}-\left(  1-i\sqrt{3}\right)  \frac{A}{2^{4/3}a}-\left(
1+i\sqrt{3}\right)  \frac{1+ac}{2^{2/3}aA},\nonumber\\
M_{+}  &  =\frac{1}{a}-\left(  1+i\sqrt{3}\right)  \frac{A}{2^{4/3}a}-\left(
1-i\sqrt{3}\right)  \frac{1+ac}{2^{2/3}aA},
\end{align}
with%
\begin{equation}
A=\left[  2+3ac-3a^{2}+\sqrt{\left(  2+3ac-3a^{2}\right)  ^{2}-4(1+ac)^{3}%
}\right]  ^{1/3},
\end{equation}
and $c=g\chi\left(  \omega^{2}n_{0}\right)  ^{-1}$. To study the modifications
to the dynamics it is enough to expand each root in powers of the small
parameter $\chi$%
\begin{equation}
M_{0}\simeq\frac{1}{\tilde{\beta}_{1}}\chi^{-1},\ \ \ M_{\pm}\simeq\pm\left[
1+\frac{1}{2}\left(  \tilde{\beta}_{1}-\tilde{\alpha}_{1}\right)  \left(
\lambda\chi+\frac{1}{4}\left(  5\tilde{\beta}_{1}-\,\tilde{\alpha}_{1}\right)
\chi^{2}\right)  \right]  , \label{P2}%
\end{equation}
where $\tilde{\beta}_{1}=h_{3}n_{0}$ and $\tilde{\alpha}_{1}=g/(\omega
^{2}n_{0})$. Since the parameter $\lambda$ and the momentum $k$ appear only in
the combination $\lambda k$ it is clear that we have the symmetry property%
\begin{equation}
G^{\lambda}(\omega,k)=G^{-\lambda}(\omega,-k). \label{simprop}%
\end{equation}
This relation will be useful in the final calculation of the Green functions
$G^{\lambda}(\omega,\mathbf{r})$.

The integral in (\ref{gg})\ can be written%
\begin{equation}
G_{ik}^{ret}(\omega,\mathbf{r})=\left[  G_{ik}^{ret}\right]  _{1}%
(\omega,\mathbf{r})+\left[  G_{ik}^{ret}\right]  _{2}(\omega,\mathbf{r}%
)+\left[  G_{ik}^{ret}\right]  _{3}(\omega,\mathbf{r}).
\end{equation}
Each vector $k_{j}$ can be obtained by inserting $-i\partial_{j}$ outside the
integral. In this way we have
\begin{align}
2\left[  G_{ik}^{ret}\right]  _{1}(\omega,\mathbf{r})  &  =\delta_{ik}%
\int\frac{d^{3}k}{\left(  2\pi\right)  ^{3}}e^{i\mathbf{k}\cdot\mathbf{r}%
}\,\sum_{\lambda}G^{\lambda}(\omega,k)=-\frac{i}{r}\frac{1}{\left(
2\pi\right)  ^{2}}\delta_{ik}\int_{-\infty}^{\infty}kdk\left(  e^{ikr}\right)
\,\left[  G^{+}(\omega,k)+G^{-}(\omega,k)\right]  ,\nonumber\\
2\left[  G_{ik}^{ret}\right]  _{2}(\omega,\mathbf{r})  &  =\partial
_{i}\partial_{k}\int\frac{d^{3}k}{\left(  2\pi\right)  ^{3}}\frac
{e^{i\mathbf{k}\cdot\mathbf{r}}}{k^{2}}\,\sum_{\lambda}G^{\lambda}%
(\omega,k)=-\frac{i}{r}\frac{1}{\left(  2\pi\right)  ^{2}}(-n_{i}n_{k}%
)\int_{-\infty}^{\infty}kdk\,e^{ikr}\left[  G^{+}(\omega,k)+G^{-}%
(\omega,k)\right]  ,\nonumber\\
2\left[  G_{ik}^{ret}\right]  _{3}(\omega,\mathbf{r})  &  =\epsilon
_{irk}\partial_{r}\int\frac{d^{3}k}{\left(  2\pi\right)  ^{3}}\frac
{e^{i\mathbf{k}\cdot\mathbf{r}}}{k}\,\sum_{\lambda}G^{\lambda}(\omega
,k)=-\frac{i}{r}\frac{1}{\left(  2\pi\right)  ^{2}}i\epsilon_{ijk}n_{j}%
\int_{-\infty}^{\infty}kdk\left(  e^{ikr}\right)  \,\left(  G^{+}%
(\omega,k)-G^{-}(\omega,k)\right)  ,\nonumber\\
&
\end{align}
where $n_{i}=x_{i}/r$. Thus we arrive at%
\begin{equation}
\left[  G_{ik}^{ret}\right]  (\omega,\mathbf{r})=-\frac{i}{r}\frac{1}{\left(
2\pi\right)  ^{2}}\sum_{\lambda}\frac{1}{2}\left(  \delta_{ik}-n_{i}%
n_{k}+i\lambda\epsilon_{ipk}n_{p}\right)  \int_{-\infty}^{\infty}%
kdke^{ikr}\,G^{\lambda}(\omega,k),
\end{equation}
which allows us to identify $G^{\lambda}(\omega,\mathbf{r})$%
\begin{equation}
G^{\lambda}(\omega,\mathbf{r})=-\frac{i}{r}\frac{1}{\left(  2\pi\right)  ^{2}%
}\int_{-\infty}^{\infty}kdke^{ikr}\,G^{\lambda}(\omega,k). \label{GLAMBDA}%
\end{equation}
Let us emphasize that the factor $e^{ikr}$ forces us to close the integration
contour by a circle at infinite in the upper half complex plane, thus picking
up the poles in this region. Our next step is to perform the integrals in
(\ref{GLAMBDA}). To this end we recall that%
\begin{equation}
G^{\lambda}(\omega,k)=-\frac{3}{n_{0}^{2}\omega^{2}Qa}\frac{1}{\left(
M^{\lambda}-M_{0}\right)  \left(  M^{\lambda}-M_{+}\right)  \left(
M^{\lambda}-M_{-}\right)  }. \label{GLDET}%
\end{equation}
Our description is an effective one valid only for momenta $k<<M_{P}$.
According to Eqs. (\ref{P2}), the pole at $M_{0}^{\lambda}$ corresponds to the
momentum value%
\begin{equation}
|k_{0}|=\left\vert Qh_{3}^{-1}\right\vert \,\tilde{\xi}^{-1}.
\end{equation}
In the approximation considered here this pole can be taken at infinity and
its contribution to the integral can be neglected. The two remaining poles,
which are the ones that contribute to the integral, are located at very small
displacements with respect to $|k_{\pm}|=n_{0}\omega<<M_{P}$. In this way we
get%
\begin{equation}
G^{\lambda}(\omega,k)=\frac{3}{n_{0}^{2}\omega^{2}QaM_{0}}\frac{1}{\left(
M^{\lambda}-M_{+}\right)  \left(  M^{\lambda}-M_{-}\right)  }.
\end{equation}

Let us consider first the case of $\ G^{+}$, where we have%
\begin{equation}
G^{+}(\omega,k)=\frac{3}{QaM_{0}}\frac{1}{\left(  k-\omega n_{0}M_{+}\right)
\left(  k-\omega n_{0}M_{-}\right)  }.
\end{equation}
From the leading order expressions in (\ref{P2})\ we conclude that the pole
that contributes in this case is $\left(  \omega+i\epsilon\right)  n_{0}M_{+}
$. Thus the resulting integral is%
\begin{equation}
G^{+}(\omega,\mathbf{r})=\frac{3}{4\pi r}\frac{1}{QaM_{0}}\,\frac{2n_{0}M_{+}%
}{\left(  n_{0}M_{+}-n_{0}M_{-}\right)  }e^{i\omega n_{0}M_{+}r}.
\end{equation}
The second integral proceeds in an analogous way, but now $\left(  -\left(
\omega+i\epsilon\right)  n_{0}M_{-}\right)  $ is the pole that contributes. It
is convenient to define the refraction indices in the circular polarization
basis as%
\begin{equation}
n_{+}(\omega)=n_{0}M_{+},\;\;\;n_{-}(\omega)=-n_{0}M_{-}. \label{REFIND}%
\end{equation}
The minus sign in $n_{-}$ is due to the fact that $M_{-}$ starts with a $-1$.
Up to the order considered, they are given by the explicit expressions%
\begin{equation}
n_{\lambda}(\omega)=n_{0}\left[  1+\lambda\left(  \tilde{\beta}_{1}%
-\tilde{\alpha}_{1}\right)  \frac{\chi}{2}+\left(  \tilde{\beta}_{1}%
-\tilde{\alpha}_{1}\right)  \left(  5\tilde{\beta}_{1}-\tilde{\alpha}%
_{1}\right)  \frac{\chi^{2}}{8}\right]  .
\end{equation}
From here we can obtain an explicit second order power expansion in $\xi$%
\begin{equation}
n_{\lambda}(\omega)=1+\lambda n_{1}\omega\tilde{\xi}+n_{2}\left(  \omega
\tilde{\xi}\right)  ^{2}+O\left(  \tilde{\xi}^{3}\right)  .
\end{equation}
where%
\begin{align}
n_{1}  &  =\frac{1}{2}\left(  \alpha_{10}-\beta_{10}\right) \\
n_{2}  &  =\frac{1}{8}\left[  \left(  \alpha_{10}-\beta_{10}\right)  \left(
\alpha_{10}-5\beta_{10}\right)  +4\left(  \beta_{02}-\alpha_{02}\right)
\right]
\end{align}
With this notation we finally get
\begin{equation}
G^{\lambda}(\omega,\mathbf{r})=\frac{1}{4\pi rQ}\frac{2n_{\lambda}}{\left(
n_{-}+n_{+}\right)  }e^{i\omega n_{\lambda}r}, \label{glor}%
\end{equation}
where we have considered that the dominant term in $M_{0}$ yields $aM_{0}=1$.
The reality conditions (\ref{RC1}-\ref{RC2}) impose the relations%
\begin{equation}
\left[  G^{\lambda}(\omega,\mathbf{r})\right]  ^{\ast}=G^{-\lambda}%
(-\omega,\mathbf{r}),\;\;\;\;\left[  n_{\lambda}(\omega)\right]  ^{\ast
}=n_{-\lambda}(-\omega).
\end{equation}
The low energy effective character of the theory implies the existence of a
frequency cutoff $\Omega$. Taking this into account, the space-time Green
function is%
\begin{align}
G^{\lambda}(\tau,\mathbf{r})  &  =\frac{1}{4\pi r}\int_{-\Omega}^{\Omega
}d\omega\frac{2n_{\lambda}}{Q\left(  n_{-}+n_{+}\right)  }e^{i\omega
n_{\lambda}r}e^{-i\omega\tau}\nonumber\\
&  =\frac{1}{4\pi r}\int_{-\Omega}^{\Omega}d\omega\left[  1+\lambda
n_{1}\omega\tilde{\xi}-\left(  \alpha_{20}-\beta_{02}\right)  \left(
\omega\tilde{\xi}\right)  ^{2}\right]  e^{i\omega\left[  1+\lambda n_{1}%
\omega\tilde{\xi}+n_{2}\left(  \omega\tilde{\xi}\right)  ^{2}\right]
r}e^{-i\omega\tau}%
\end{align}
where $\tau=t-t^{\prime}$. The choice of the poles in the complex plane
$\omega$ is consistent with the causal behavior of the Green function. But the
frequency cutoff could introduce some violation of causality. To investigate
this\textbf{\ }possibility, we can compute the time dependent Green function
by expanding the integrand in powers of ${\tilde{\xi}}$%
\begin{align}
G^{\lambda}(\tau,\mathbf{r})  &  \simeq\frac{1}{4\pi r}\int_{-\Omega}^{\Omega
}d\omega\left\{  1+\lambda n_{1}\left(  1+i\omega r\right)  \omega{\tilde{\xi
}}-\left[  \alpha_{20}-\beta_{02}-i\left(  n_{2}+n_{1}^{2}\right)
r\omega+\frac{1}{2}n_{1}^{2}r^{2}\omega^{2}\right]  \omega^{2}{\tilde{\xi}%
}^{2}\right\}  e^{i\omega\left(  r-\tau\right)  }\nonumber\\
&  =\frac{1}{2\pi r}\left\{  1-i\lambda n_{1}{\tilde{\xi}}\left(
1+r\partial_{r}\right)  \partial_{r}+{\tilde{\xi}}^{2}\left[  \alpha
_{20}-\beta_{02}-\left(  n_{2}+n_{1}^{2}\right)  r\partial_{r}-\frac{1}%
{2}n_{1}^{2}r^{2}\partial_{r}^{2}\right]  \partial_{r}^{2}\right\}  \frac
{\sin\left(  r-\tau\right)  \Omega}{r-\tau}.
\end{align}
This expression shows that the main effect of the cutoff is to spread the
propagating field around the light cone, within a wedge defined by
$r\simeq\tau\pm\pi/2\Omega$. Returning to $G^{\lambda}(\omega,\mathbf{r})$,
Eq. (\ref{glor}), we can characterize the effect of the cutoff in the causal
behavior of the Green function using the generalized susceptibility theorem
\cite{LANDAU1}, a generalization of the Kramers-Kronig relations. In terms of
the frequency $\omega$, the real and imaginary parts of the Green function are%
\begin{align}
\text{Re}\;G^{\lambda}(\omega,r)  &  \simeq\frac{1}{4\pi r}\left\{  \left[
1+{\tilde{\xi}}^{2}\left(  \alpha_{20}-\beta_{02}-\left(  n_{2}+n_{1}%
^{2}\right)  r\partial_{r}-\frac{1}{2}n_{1}^{2}r^{2}\partial_{r}^{2}\right)
\partial_{r}^{2}\right]  \cos\omega r+\lambda n_{1}{\tilde{\xi}}\left(
1+r\partial_{r}\right)  \partial_{r}\sin\left(  \omega r\right)  \right\}
,\nonumber\\
& \label{reg}\\
\text{Im}\;G^{\lambda}(\omega,r)  &  \simeq\frac{1}{4\pi r}\left\{  \left[
1+{\tilde{\xi}}^{2}\left(  \alpha_{20}-\beta_{02}-\left(  n_{2}+n_{1}%
^{2}\right)  r\partial_{r}-\frac{1}{2}n_{1}^{2}r^{2}\partial_{r}^{2}\right)
\partial_{r}^{2}\right]  \sin\omega r-\lambda n_{1}{\tilde{\xi}}\left(
1+r\partial_{r}\right)  \partial_{r}\cos\omega r\right\}  .\nonumber\\
&  \label{img}%
\end{align}
According to the generalized susceptibility theorem the Green function will be
causal if
\begin{equation}
\text{Im\ }G(\omega)=-\frac{1}{\pi}P\int_{-\Omega}^{\Omega}d\omega^{\prime
}\frac{\text{Re}\;G(\omega^{\prime})-\text{Re}\;G(\Omega)}{\omega^{\prime
}-\omega},
\end{equation}
or more explicitly when%
\begin{align}
\text{Im}G^{\lambda}(\omega\text{)}  &  =-\frac{1}{4\pi^{2}r}\left\{  \left[
1+{\tilde{\xi}}^{2}\left(  \alpha_{20}-\beta_{02}-\left(  n_{2}+n_{1}%
^{2}\right)  r\partial_{r}-\frac{1}{2}n_{1}^{2}r^{2}\partial_{r}^{2}\right)
\partial_{r}^{2}\right]  \ P\int_{-\Omega}^{\Omega}d\omega^{\prime}\frac
{\cos\left(  \omega^{\prime}r\right)  -\cos\left(  \Omega r\right)  }%
{\omega^{\prime}-\omega}\right. \nonumber\\
&  \left.  +\lambda n_{1}{\tilde{\xi}}\left(  1+r\partial_{r}\right)
\partial_{r}\ P\int_{-\Omega}^{\Omega}d\omega^{\prime}\frac{\sin\left(
\omega^{\prime}r\right)  -\sin\left(  \Omega r\right)  }{\omega^{\prime
}-\omega}\right\}  . \label{imgkk}%
\end{align}
For $\omega/\Omega\ll1$ the integrals reduce to%
\begin{align}
P\int_{-\Omega}^{\Omega}d\omega^{\prime}\frac{\cos\left(  \omega^{\prime
}r\right)  -\cos\left(  \Omega r\right)  }{\omega^{\prime}-\omega}  &
\simeq2\left(  1-\cos\omega R\right)  \frac{\omega}{\Omega}\cos\left(  \Omega
r\right)  -\left[  \pi+\left[  \left(  \Omega r\right)  \left(  \cos\Omega
r\right)  -\sin\Omega r\right]  \left(  \frac{\omega}{\Omega}\right)
^{2}\right]  \sin\omega r,\nonumber\\
P\int_{-\Omega}^{\Omega}d\omega^{\prime}\frac{\sin\left(  \omega^{\prime
}r\right)  -\sin\left(  \Omega r\right)  }{\omega^{\prime}-\omega}  &
\simeq2\left(  1-\sin\omega r\right)  \frac{\omega}{\Omega}\cos\left(  \Omega
r\right)  +\left[  \pi+\left[  \left(  \Omega r\right)  \left(  \cos\Omega
r\right)  -\sin\Omega r\right]  \left(  \frac{\omega}{\Omega}\right)
^{2}\right]  \cos\omega r.\nonumber\\
&
\end{align}
Furthermore, in the case of a radiation field $\Omega r\gg1$ and hence the
factors $\cos\Omega r$ and $\sin\Omega r$ become strongly oscillating,
nullifying the contributions of the terms where they appear (which also have a
factor $\left(  \omega/\Omega\right)  ^{n}$, with $n\geq1$). Thus we can take%
\begin{equation}
P\int_{-\Omega}^{\Omega}d\omega^{\prime}\frac{\cos\left(  \omega^{\prime
}r\right)  -\cos\left(  \Omega r\right)  }{\omega^{\prime}-\omega}\simeq
-\pi\sin\omega r,\qquad P\int_{-\Omega}^{\Omega}d\omega^{\prime}\frac
{\sin\left(  \omega^{\prime}r\right)  -\sin\left(  \Omega r\right)  }%
{\omega^{\prime}-\omega}\simeq\pi\cos\omega r.
\end{equation}
Replacing these integrals in (\ref{imgkk}), we finally get that if Re
$G^{\lambda}(\omega)$ is given by Eq. (\ref{reg}), the imaginary part of the
Green function, Im $G^{\lambda}(\omega)$, must be
\begin{equation}
\text{Im}\;G^{\lambda}(\omega)=\frac{1}{4\pi r}\left\{  \left[  1+{\tilde{\xi
}}^{2}\left(  \alpha_{20}-\beta_{02}-\left(  n_{2}+n_{1}^{2}\right)
r\partial_{r}-\frac{1}{2}n_{1}^{2}r^{2}\partial_{r}^{2}\right)  \partial
_{r}^{2}\right]  \sin\omega r-\lambda n_{1}{\tilde{\xi}}\left(  1+r\partial
_{r}\right)  \partial_{r}\cos\omega r\right\}  ,
\end{equation}
which coincides with Eq. (\ref{img}), obtained by direct computation. This
result shows that the cutoff does not introduce any significant causality
violation provided that $\Omega r>>1$ and $\omega/\Omega<<1$.

\section{Final comments}

In the preceding sections we have discussed the implications of reality and
causality of the electromagnetic fields upon the structure of the constitutive
relations and the Green functions in quantum gravity inspired modified
electrodynamics. To illustrate these implications we now apply our results to
some particular models already found in the literature.

Let us start with the Gambini-Pullin electrodynamics \cite{GP}, where the
constitutive relations are
\begin{equation}
\mathbf{D}=\mathbf{E}-2i{\tilde{\xi}\omega}\mathbf{B}+4{\tilde{\xi}}^{2}%
\omega\;\mathbf{k\times B,\;\;\;H}=\mathbf{B}+2i{\tilde{\xi}}\mathbf{k}%
\times\mathbf{B}. \label{GPCR}%
\end{equation}
Introducing the dominant contributions to satisfy causality they become%
\begin{equation}
\mathbf{D}=\mathbf{E}-2i{\tilde{\xi}\omega}\mathbf{B}+4\left(  1+\frac
{16}{3\pi}\frac{i\omega}{\Omega}\right)  {\tilde{\xi}}^{2}\omega
\;\mathbf{k\times B,\;\;\;H}=\mathbf{B}+2i{\tilde{\xi}}\mathbf{k}%
\times\mathbf{B},
\end{equation}
which leads to the refraction index
\begin{equation}
n_{\lambda}(\omega)\simeq1+2\lambda\omega{\tilde{\xi}}+4\omega^{2}{\tilde{\xi
}}^{2}-i\frac{64}{3\pi}\frac{\omega}{\Omega}\omega^{2}{\tilde{\xi}}^{2},
\end{equation}
indicating that absorption is generated at the order $({\tilde{\xi}}%
\omega)^{2}$.

The second case is the model of Myers and Pospelov \cite{MP}, which yields
\begin{equation}
\mathbf{D}=\mathbf{E}+i\tilde{\xi}\omega\mathbf{B,\;\;\;\;\;H}=\mathbf{B}%
+i\tilde{\xi}\omega\mathbf{E}.
\end{equation}
Here the constitutive relations become%
\begin{equation}
\mathbf{D}=\mathbf{E}+i\left(  1+\frac{\omega^{2}}{\Omega^{2}}\right)
\tilde{\xi}\omega\mathbf{B,\;\;\;\;\;\;H}=\mathbf{B}+i\left(  1+\frac
{\omega^{2}}{\Omega^{2}}\right)  \tilde{\xi}\omega\mathbf{E},
\end{equation}
after the causality requirement is imposed, and the resultant index of
refraction is
\begin{equation}
n_{\lambda}(\omega)\simeq1-\lambda\left(  1+\frac{\omega^{2}}{\Omega^{2}%
}\right)  \omega\tilde{\xi}+\frac{1}{2}\left(  1+\frac{\omega^{2}}{\Omega^{2}%
}\right)  ^{2}\omega^{2}\tilde{\xi}^{2}.
\end{equation}
In this case the absorptive terms begin at least to order $\omega^{5}$ and
causality only introduces $\Omega$-dependent corrections to the velocity of
propagation. The two models discussed previously present birefringence.

The third model that we consider is the flat space version of that of Ellis et
al.\cite{EDMED2,ELLIS2,ELLIS3}, which has been used to study in a rough way
the phenomenology of the full underlying model, leading to robust limits on
Lorentz violation. From the theoretical model of a recoiling $D$-particle in a
quantum gravitational foam, a simple effective model formally analogous to
electrodynamics in a medium has been heuristically introduced in Ref.
\cite{ELLIS2}. Such an approximation has been used in conjunction with
observations of gamma ray bursts to set limits on the quantum gravity scale
$M$\cite{ELLIS3}. This effective model is characterized by constitutive
relations\cite{ELLIS2}%
\begin{equation}
\mathbf{D}=\frac{\mathbf{E}}{\sqrt{h}}+\mathbf{H}\times\mathbf{G}%
,\;\;\;\;\;\;\;\;\mathbf{B}=\frac{\mathbf{H}}{\sqrt{h}}-\mathbf{E}%
\times\mathbf{G},
\end{equation}
where $1/\sqrt{h}$ plays the role of the electric and magnetic permeability
and is taken equal to one to have the same permeability as the classical
vacuum. In references \cite{ELLIS2,GRAVANIS} the vector $\mathbf{G}$, which
originally represents the recoil velocity of the $D$-particle (being
proportional to the momentum transfer over $M$)\ is very crudely identified
with $\mathbf{G}\sim\mathbf{k}/M$, where $\mathbf{k}$ is the momentum of the
propagating photon and $M$ is the effective mass of the $D$-particle in the
quantum space-time foam. In our notation this amounts to require%
\[
\mathbf{G=}f\;\mathbf{k}%
\]
and the above constitutive relations turn out to be
\begin{equation}
\mathbf{D}=\mathbf{E}+\left(  f^{2}\omega-f\right)  \left(  \mathbf{k}%
\times\mathbf{B}\right)  ,\;\;\mathbf{H}=\left(  1-f\omega\right)  \mathbf{B}.
\end{equation}
To lowest order, the reality conditions for the fields imply that we must
choose%
\begin{equation}
f=ia\tilde{\xi},
\end{equation}
with $a$ being a constant real number. Imposing causality we get%
\begin{equation}
\mathbf{D}=\mathbf{E}+ia\left[  \left(  1+\frac{\omega^{2}}{\Omega^{2}%
}\right)  -ia\omega\tilde{\xi}\right]  \tilde{\xi}\left(  \mathbf{k}%
\times\mathbf{B}\right)  ,\ \ \ \ \ \ \ \ \ \ \mathbf{H}=\left[  1-ia\left(
1+\frac{\omega^{2}}{\Omega^{2}}\right)  \tilde{\xi}\omega\right]  \mathbf{B}.
\end{equation}
The dominant contributions to the refraction index do not depend on the causal
corrections and are
\begin{equation}
n_{\lambda}(\omega)=1+ia\omega\tilde{\xi}-a^{2}\omega^{2}\tilde{\xi}^{2}.
\end{equation}
Contrary to the previous cases, the above indicates that absorption is already
present to linear order (a point previously missed) and that the propagation
properties of this effective model exhibit corrections starting from second
order in ${\tilde{\xi}}$. In fact the photon group velocity is given by
\begin{equation}
c(\omega)=\frac{d\,(\mathrm{Re}\,k)}{d\omega}=1-3(a{\tilde{\xi}}\omega)^{2},
\end{equation}
and not by $c(\omega)=c_{0}(1-\omega/M)$, with $M$ real, as was originally
assumed in setting the observational bounds \cite{EDMED2}.

To summarize, our discussion is based on the consideration that these modified
electrodynamics are effective theories with physical meaning, and hence that
in particular the electromagnetic field strengths $\mathbf{E}$ and
$\mathbf{B}$, and the excitations $\mathbf{D}$ and $\mathbf{H}$, must be real
and its propagation causal. We have shown that these requirements impose
constraints on the structure of the constitutive relations and lead to
additional contributions to the photon propagation velocity and, much more
interesting, to absorptive effects in the propagation of the fields. The
contributions originating from causality depend on the range of validity of
the effective theory, coded here by the cutoff $\Omega$.

\section*{Acknowledgements}

R.M. acknowledges partial support from CONICET-Argentina. L.F.U is partially
supported by projects CONACYT-40745F and DGAPA-UNAM-IN104503-3. R.M and L.F.U
thank H. Vucetich for very useful discussions. L.F.U. thanks N. Mavromatos for
comments and clarifications.

\section*{Appendix A}

In this Appendix we analyze the effect of a frequency cutoff on the causal
character of the derivative of a causal function $f(t)$, defined by a
generalized susceptibility $\vartheta(t)$%
\begin{equation}
f(t)=\int_{-\infty}^{\infty}d\tau\ \vartheta(t-\tau)g\left(  \tau\right)  .
\end{equation}
The derivative is%
\begin{equation}
\partial_{t}f(t)=\int_{-\infty}^{\infty}d\tau\ \partial_{t}\vartheta
(t-\tau)g\left(  \tau\right)  ,
\end{equation}
and therefore the problem reduces to studying the causal character of the
derivative of $\vartheta(t)$, considered itself as a susceptibility function.
We are dealing with real functions, so we will consider here a real
susceptibility , $\vartheta(t)=\vartheta^{\ast}(t)$. Introducing the Fourier
transform%
\begin{equation}
\vartheta(\omega)=\int_{-\infty}^{\infty}dt\ e^{i\omega t}\vartheta(t),
\end{equation}
we get in this case
\begin{equation}
\mathrm{Re}\vartheta(\omega)=\mathrm{Re}\vartheta(-\omega
),\ \ \ \ \ \mathrm{Im}\vartheta(\omega)=-\mathrm{Im}\vartheta(-\omega).
\end{equation}
If $\vartheta(t)$ is causal, its Fourier transform satisfies the
Kramers-Kronig relations, which in the presence of a frequency cutoff $\Omega$
become%
\begin{align}
\mathrm{\mathrm{Re}}\,\vartheta(\omega)  &  =\frac{1}{\pi}P\int_{-\Omega
}^{\Omega}d\omega^{\prime}\ \frac{\mathrm{\mathrm{Im}}\,\vartheta
(\omega^{\prime})}{\omega^{\prime}-\omega}+\kappa,\label{pp11}\\
\mathrm{\mathrm{Im}}\,\vartheta(\omega)  &  =-\frac{1}{\pi}P\int_{-\Omega
}^{\Omega}d\omega^{\prime}\ \frac{\mathrm{\mathrm{Re}}\,\vartheta
(\omega^{\prime})-\kappa}{\omega^{\prime}-\omega}+\frac{A}{\omega^{m}}.
\label{pp22}%
\end{align}
where $\kappa=\mathrm{\mathrm{\mathrm{Re}}}\vartheta(\Omega)$ and it is
assumed that $\vartheta(\omega)=i\frac{A}{\omega^{m}}$ at $\omega\simeq0 $.
The Fourier transform of the derivative of the susceptibility function is
given by%
\begin{equation}
\tilde{\vartheta}(\omega)=-i\omega\vartheta(\omega),
\end{equation}
and hence
\begin{align}
\mathrm{\mathrm{Re}}\,\tilde{\vartheta}(\omega)  &  =\omega\mathrm{\mathrm{Im}%
}\,\vartheta(\omega),\label{re}\\
\mathrm{\mathrm{\mathrm{Im}}}\,\tilde{\vartheta}(\omega)  &  =-\omega
\mathrm{Re}\,\vartheta(\omega). \label{im}%
\end{align}
In order for $\tilde{\vartheta}(\omega)$ to also be causal, it must satisfy
the corresponding Kramers-Kronig relations
\begin{align}
\mathrm{Re}\tilde{\vartheta}(\omega)  &  =\frac{1}{\pi}P\int_{-\Omega}%
^{\Omega}d\omega^{\prime}\ \frac{\mathrm{Im}\,\tilde{\vartheta}(\omega
^{\prime})}{\omega^{\prime}-\omega}+\tilde{\kappa},\\
\mathrm{Im}\,\tilde{\vartheta}(\omega)  &  =-\frac{1}{\pi}P\int_{-\Omega
}^{\Omega}d\omega^{\prime}\ \frac{\mathrm{Re}\,\tilde{\vartheta}%
(\omega^{\prime})-\tilde{\kappa}}{\omega^{\prime}-\omega}+\frac{\tilde{A}%
}{\omega^{n}},
\end{align}
where $\tilde{\kappa}=\mathrm{Re}\tilde{\vartheta}(\Omega)=\left.
\omega\mathrm{Im}\,\vartheta(\omega)\right\vert _{\Omega}$ and we are assuming
that $\tilde{\vartheta}(\Omega)$ has a pole of order $n$ at $\omega=0$.
Rewriting $\mathrm{Re}\tilde{\vartheta}(\omega)$ and $\mathrm{Im}%
\,\tilde{\vartheta}(\omega)$ in terms of $\mathrm{Re}\,\vartheta(\omega)$ and
$\mathrm{Im}\,\vartheta(\omega)$, according to (\ref{re}) and (\ref{im}),
these last equations yield%
\begin{align}
\mathrm{Im}\,\vartheta(\omega)  &  =-\frac{1}{\pi\omega}P\int_{-\Omega
}^{\Omega}d\omega^{\prime}\ \frac{\omega^{\prime}\mathrm{Re}\,\vartheta
(\omega^{\prime})}{\omega^{\prime}-\omega}+\frac{\tilde{\kappa}}{\omega
},\label{qq1}\\
\mathrm{Re}\,\vartheta(\omega)  &  =\frac{1}{\pi\omega}P\int_{-\Omega}%
^{\Omega}d\omega^{\prime}\ \frac{\omega^{\prime}\mathrm{Im}\,\vartheta
(\omega^{\prime})-\tilde{\kappa}}{\omega^{\prime}-\omega}-\frac{\tilde{A}%
}{\omega^{n+1}}. \label{qq2}%
\end{align}

Finally, comparing Eqs. (\ref{pp11}-\ref{pp22}) and (\ref{qq1}-\ref{qq2}), and
considering that $\mathrm{Im}\,\vartheta(\omega)$ is an odd function, we get%
\begin{align}
\kappa+\frac{\tilde{A}}{\omega^{n+1}}+\frac{\tilde{\kappa}}{\pi\omega}%
\ln\left(  \frac{1-\omega/\Omega}{1+\omega/\Omega}\right)   &  =0,\\
\frac{1}{\pi\omega}\int_{-\Omega}^{\Omega}d\omega^{\prime}\ \mathrm{Re}%
\,\vartheta(\omega^{\prime})  &  =-\frac{\kappa}{\pi}\ln\left(  \frac
{1-\omega/\Omega}{1+\omega/\Omega}\right)  -\frac{A}{\omega^{m}}+\frac
{\tilde{\kappa}}{\omega}.
\end{align}
\qquad The first relation implies%
\begin{equation}
\tilde{A}=\kappa=\tilde{\kappa}=0, \label{nc1}%
\end{equation}
and thus the second one leads to%
\begin{equation}
m=1,\ \ \ \ \ \ \ \ \ \ \ \ \ \int_{-\Omega}^{\Omega}d\omega^{\prime
}\ \mathrm{Re}\,\vartheta(\omega^{\prime})=-\pi A. \label{nc3}%
\end{equation}
When there is no cutoff at a finite frequency the factor $\ln\left(
\frac{1-\omega/\Omega}{1+\omega/\Omega}\right) $ becomes $0$, and the
constraints are%
\begin{align}
\kappa &  =\tilde{A}=0,\label{cc1}\\
\int_{-\infty}^{\infty}d\omega^{\prime}\ \mathrm{Re}\,\vartheta(\omega
^{\prime})  &  =\pi\left(  \tilde{\kappa}-\frac{A}{\omega^{m-1}}\right)  .
\label{cc2}%
\end{align}
For example, let us consider what happens when $\vartheta(t)$ is the causal
Heaviside function, such that $\vartheta(\omega)=i/\omega$ and $\tilde
{\vartheta}(\omega)=1$. In both cases, $\Omega=\infty$ or $\Omega$ finite, we
get%
\begin{equation}
\kappa=\tilde{A}=0\ \ \ \ \ \ \ A=\tilde{\kappa}=1,\ \ \ \ \int_{-\Omega
}^{\Omega}d\omega^{\prime}\ \mathrm{Re}\,\vartheta(\omega^{\prime})=0.
\end{equation}
When $\Omega=\infty$ the constraints (\ref{cc1}-\ref{cc2}) hold and are indeed
satisfied, and hence $\partial_{t}\theta(t)=\delta(t)$ is also a causal
susceptibility. But if there is a finite frequency cutoff $\Omega$ the
corresponding constraints are given by (\ref{nc1}-\ref{nc3}), which are not
satisfied, and $\delta(t)$ becomes a non causal susceptibility.

In the case of a field theory we have spatial derivatives together with the
time derivative. But the former derivatives are related to the latter by the
equations of motion that drive the evolution of the system. Thus, to discuss
causal characteristics it is necessary to take into account both types of
derivatives, linked at the Fourier transform level by the refraction index
that relates frequency and momentum.

\end{document}